\documentclass[conference]{IEEEtran} % Use 'journal' for journal submissions
\usepackage{amsmath} % For math symbols
\usepackage{graphicx} % For figures
\usepackage{cite} % For IEEE-style citations
\usepackage{hyperref} % Optional, for clickable references
\newcommand{\Sig}[1]{\textrm{S} \{ #1\} }

\title{Phase Alignment Enhances Oscillatory Power in Neural Mass Models Optimized for Class Encoding}
\author{
    \IEEEauthorblockN{Alexander Pei}
    \IEEEauthorblockA{
        Email: apei2@jh.edu
    }
}

\begin{document}
\maketitle

\begin{abstract} Neural encoding of objects and cognitive states remains an elusive yet crucial aspect of brain function. While traditional feed-forward machine learning neural networks have enormous potential to encode information, modern architectures provide little insight into the brain's mechanisms. In this work, a Jansen and Rit neural mass model was constructed to encode different sets of inputs, aiming to understand how simple neural circuits can represent information. A genetic algorithm was used to optimize parameters that maximized the differences in responses to particular inputs. These differences in responses manifested as phase-shifted oscillations across the set of inputs. By delivering impulses of excitation synchronized with a particular phase-shifted oscillation, we demonstrated that the encoded phase could be decoded by measuring oscillatory power. These findings demonstrate the capability of neural dynamical circuits to encode and decode information through phase. \end{abstract}

% \begin{IEEEkeywords} % Neural encoding, phase shifts, Jansen and Rit neural mass model, genetic algorithms, oscillatory power % \end{IEEEkeywords}

\section{Introduction} In 2005, researchers announced the discovery of the "Jennifer Aniston neuron"—a neuron that selectively fired in response to images of the actress Jennifer Aniston \cite{quiroga2005invariant}. While this finding sparked considerable debate, it exemplifies how the brain may encode complex, high-level concepts within neural circuitry. However, the mechanisms underpinning how the brain encodes such information remain unknown. Although single neurons may selectively respond to particular stimuli, they never function in isolation. Each neuron is influenced by thousands of direct presynaptic inputs and the extracellular environment in which it exists. Imaging modalities such as electroencephalography (EEG) and functional magnetic resonance imaging (fMRI), which capture the collective activity of hundreds of thousands of neurons, reveal distinct patterns of activity associated with specific cognitive states. This raises the question of how the coordinated action of large neural networks gives rise to the brain's ability to encode increasingly abstract concepts. The aim of this work is to study how biologically inspired neural dynamical systems encode and decode information.

To address this challenge, we turn to mesoscopic computational models, which represent the activity of thousands of neurons through simplified equations. Modeling each individual neuron among the millions involved would provide the highest-fidelity computational model but is intractable due to computational constraints. One such mesoscopic model is the Jansen and Rit neural mass model (JRNMM), which describes the collective behavior of thousands of cortical neurons \cite{Jansen1995}. This model consists of a simple cortical circuit of interconnected excitatory interneurons, inhibitory interneurons, and pyramidal cells to simulate cortical dynamics. The JRNMM has seen considerable application across various neuroscientific disciplines. In the context of epilepsy modeling, altering synaptic parameters in the JRNMM can result in seizure-like activity when the model bifurcates into a limit cycle \cite{Wendling2000}. Although this example does not reflect traditional encoding and decoding of neural information, it demonstrates that varying synaptic parameters leads to vastly different neural dynamics. The JRNMM has also been a valuable tool in studying neural circuitry associated with specific cognitive states within the dynamic causal modeling (DCM) framework. In DCM, observed EEG data from different cognitive states are modeled by altering the parameters of the JRNMM \cite{kiebel2008dynamic}. While DCM can model particular cognitive states, it does not explicitly model the mechanisms through which the parameters change. That is, the differences in the modeled neural responses are a result of two neural dynamical systems with distinct parameters, rather than a single dynamical system responding uniquely to distinct inputs, as seen in traditional feed-forward neural networks.

In this work, we explore the ability of the JRNMM to encode neural inputs. By leveraging a genetic algorithm to maximize the differences in the JRNMM's output in response to a set of neural inputs, the model learned to represent different inputs as oscillations with distinct phase shifts. This finding aligns with the concept of phase encoding of neural information reported in existing literature. To decode the neural inputs from the phase-shifted oscillations, we delivered spikes either in phase or out of phase with the encoded oscillations. When the spikes were phase-aligned, the oscillatory power was stronger, demonstrating successful decoding of the encoded information.

\section{Background}

\subsection{Jansen and Rit Neural Mass Model}

The Jansen and Rit Neural Mass Model (JRNMM) consists of three cortical neural populations: excitatory interneurons (EI), inhibitory interneurons (II), and pyramidal cells (PC). These neuron populations are assumed to be the main drivers of the dynamics of the six layers in the cortex and comprise a small cortical column of approximately 20–200~\(\mu\)m in radius \cite{DeFelipe2002}.

Spikes from one neuron serve as inputs to another neuron by causing excitatory or inhibitory postsynaptic potentials (EPSPs and IPSPs). The total number of spikes that arrive at a neuron is determined by the average pulse density of the presynaptic neuron, multiplied by the number of synapses made onto the postsynaptic neuron. The resulting pulse density is then convolved with the postsynaptic potential (PSP) response functions.

The net sum of the PSPs in a neuron polarizes the membrane. Using Freeman's derivation of the sigmoidal wave-to-pulse function, the membrane polarization of a neuron can be converted to an average number of spikes per second to serve as the presynaptic input to another neuron population \cite{Freeman1975, Freeman1987}. The input-output process described above for a single neuron is assumed to be equivalent for the rest of the population to which that neuron belongs (e.g., neuron 1 and neuron 2 in the PC population have identical excitation and firing rate outputs).

The EI and II populations synapse onto the PC population. When an EI or II fires, it results in depolarization or hyperpolarization of the PC membrane potential, as specified by the PSP activation functions, or kernels, \(h_e(t)\) and \(h_i(t)\). This polarization is scaled by the pulse density of the neuron populations, as well as the number of synapses formed onto the PC population. The resulting polarization of the PC membrane potential is converted to a pulse density via the sigmoid function. The spiking of the PC population then provides excitatory input to the EI and II populations, depolarizing their membrane potentials via the same EPSP kernel \(h_e(t)\). The equations for the system are presented below and describe the membrane polarization voltage \(V(t)\) of each neural population:

\begin{align} \label{eqn:JR_ConvSingleCol}
    V_1(t) &= h_e(t) \ast \left[ c_1 \Sig{V_3(t)} + P(t) \right] \\
    V_2(t) &= h_e(t) \ast c_3 \Sig{V_3(t)} \\
    \begin{split}
    V_3(t) &= h_e(t) \ast c_2 \Sig{V_1(t)}  -h_i(t) \ast c_4 \Sig{V_2(t)} 
    \end{split}
\end{align}

\noindent where \(V_1(t)\), \(V_2(t)\), and \(V_3(t)\) are the membrane potentials of the EI, II, and PC populations, respectively. \(P(t)\) models the external inputs (of thalamic origin or from another cortical column) exciting the EI layer. The symbol \(\ast\) denotes the convolution operator, and \(\Sig{\cdot}\) is the sigmoid function:

\begin{equation} \label{eqn:JR_Sigmoid}
    \Sig{V(t)} = \frac{2e_0}{ 1 + \exp(r(V_0 - V(t))) }
\end{equation}

The PSP kernels are:

\begin{equation} \label{eqn:JR_EPSP}
    h_e(t) = 
    \begin{cases}
        Aat \exp(-at) & \text{if } t \geq 0, \\
        0 & \text{if } t < 0.
    \end{cases}
\end{equation}

\begin{equation} \label{eqn:JR_IPSP}
    h_i(t) = 
    \begin{cases}
        Bbt \exp(-bt) & t \geq 0 \\
        0 & t < 0
    \end{cases}
\end{equation}

% \begin{figure}[t]
%     \centering
%     \includegraphics[width=0.8\columnwidth]{figs/fig_PSP_Sigmoid.png}
%     \caption{(a) The excitatory and inhibitory PSP kernels. The points $(1/b, B/e)$ and $(1/a, A/e)$ are the peak values of the kernels. $e$ is Euler's mathematical constant, not a parameter. Shorter time constants $a$ and $b$ result in faster kernel dynamics. The peaks of the kernels are proportional to $A$ and $B$. (b) The sigmoid function $\Sig{\cdot}$ transforms the average membrane potential (mV) into a mean firing rate (action potentials per second). The maximum firing rate is $2e_0$. The point $(V_0, e_0)$ is when the population reaches half of the maximum firing rate. The slope of the function evaluated at this point is $e_0r/2$. Increasing $r$ increases the slope around this point, resulting in sharper transitions from inactivity to activity in the neural population.}
%     \label{fig:JRNMM_PSPandSigmoid}
% \end{figure}

\begin{table}[h]
\footnotesize
\caption{JRNMM parameters}
\begin{center}
\scalebox{1}{
\begin{tabular}{c c c}
\hline
\textbf{Variable}&\textbf{Description}&\textbf{Value}\\
\hline
$A, B$ & Max amplitude of EPSP and IPSP & 3.25, 29.3 (mV)\\
$a, b$ & Lumped time constants of PSP delays  & 100, 66.7 ($\text{s}^{-1}$)\\
$e_0$ & Max firing rate of a neuron & 2.5 ($\text{s}^{-1}$)\\
$V_0$ & Mean firing threshold & 0 (mV)\\
$r$ & Steepness of the sigmoid function  & 0.56  ($\text{mV}^{-1}$) \\
$c_1$ & No. of synapses a EI gets from PCs & 50 \\
$c_2$ & No. of synapses a PC gets from EIs & 40 \\
$c_3$ & No. of synapses a II gets from PCs & 12 \\
$c_4$ & No. of synapses a PC gets from IIs & 12 \\

\hline
\end{tabular}
\label{tab:JR_Params}
}
\end{center}
\end{table}

Tab.~\ref{tab:JR_Params} contains a table of the parameters. The variables $A$ and $B$ determine the maxmimum amplitude of the EPSPs and IPSPs. $a$ and $b$ are the time constants for the PSP exponential function, and determine the temporal dynamics of the PSPs. The time constants are referred as a lumped parameter representation of all the synaptic and dendritic delays, i.e. how long it takes for the presynaptic activity to manifest as a PSP near the axon hillock of the postsynaptic neuron (due to delays from propagation down the axial length of the neuron, neurotransmitter release, and other delays). 

\subsection{Multiple column extension}

The single-column JRNMM can be extended to model intercortical connectivity. A large majority of connections between cortical columns are formed by PCs in the supragranular and infragranular layers \cite{felleman1991distributed}. Intercortical interactions can be divided into three categories depending on where the axons of the PCs terminate. In ``feedforward" connections, axons of the PCs terminate in layer 4. In ``feedback" connections, axons of the PCs avoid layer 4. In ``lateral" connections, axons of the PCs terminate uniformly across the cortical column \cite{felleman1991distributed}. Applying these connectivity rules results in the following: the membrane potential of the PCs from one column is converted into a firing rate via the sigmoid function to target different neuron populations based on the aforementioned connectivity rules \cite{David2005}. In forward connectivity, the PCs excite the EIs, while in backward connectivity, the PCs excite both the PCs and the IIs. Lateral connectivity targets all cortical layers. Extending these rules to the single-column model in Eqn.~\ref{eqn:JR_ConvSingleCol} results in additional firing rate inputs to each EI, II, and PC population's voltage, scaled by the number of postsynaptic synapses from the originating population. For example, a forward connection from column $j$ to column $i$ would result in:

\begin{equation}
    V_{1,i}(t) = h_e(t) \ast \left[ c_1 \Sig{V_3(t)} + P(t) + A_f(i,j) \Sig{V_{3,j}(t)} \right]
\end{equation}

\noindent where $V_{1,i}(t)$ is the membrane potential of the EIs in column $i$, $V_{3,j}(t)$ is the PC membrane potential of column $j$, and $A_f(i,j)$ specifies the number of synapses onto the EI layer. The full state equations can be found in \cite{David2005}.

\section{Methods}

\subsection{Model Connectivity}

The following model connectivity was studied to mimic a feed-forward perceptron/hierarchical brain structure:

\begin{itemize}
    \item Four cortical columns in the ``input layer'' (columns 1--4)
    \item Six cortical columns in the first ``hidden layer'' (columns 5--10)
    \item Five cortical columns in the second ``hidden layer'' (columns 11--15)
    \item One cortical column in the ``output layer'' (column 16)
\end{itemize}

The forward connectivity is described in Table~\ref{table:forward_connections}. Within each hidden layer, columns have lateral connections between adjacent columns (e.g., columns (5,6), (6,7), $\hdots$, (9,10) have lateral connections). A propagation delay of 20 ms was simulated for intercolumn connections between units.

\begin{table}[h]
\centering
\caption{Forward Connections}
\resizebox{0.45\linewidth}{!}{%
\begin{tabular}{|c|c|}
\hline
\textbf{From Unit(s)} & \textbf{To Unit(s)}  \\
\hline
1                & 5, 6, 7                    \\
2                & 6, 7, 8                    \\
3                & 7, 8, 9                    \\
4                & 8, 9, 10                   \\
5, 6, 7          & 11, 12                     \\
6, 7, 8          & 12, 13                     \\
7, 8, 9          & 13, 14                     \\
8, 9, 10         & 14, 15                     \\
11--15           & 16 (Output Unit)           \\
\hline
\end{tabular}%
}
\label{table:forward_connections}
\end{table}

\subsection{Model Inputs} \label{sec:modelInputs}

Four distinct inputs stimulated the dynamic model depending on the input class. The input consisted of a single impulse of mean firing rate delivered to the EI layer of the input layer, with a value of 1 action potential per second. The four classes are as follows:

\begin{itemize}
    \item Class 1: 1, 0, 1, 0
    \item Class 2: 0, 1, 0, 1
    \item Class 3: 1, 1, 0, 0
    \item Class 4: 0, 0, 1, 1
\end{itemize}

\noindent where a 1 indicates that an input spike is delivered to the $i$th unit at the respective index. For example, Class 1 inputs indicate that input unit 1 and input unit 3 received an excitatory input. Each input spike occurred at $t = 0.1$ seconds and had a duration of 0.01 seconds.

\subsection{Model Simulation and Genetic Algorithm}

A delay differential equation form of the JRNMM equations was solved using \texttt{dde23} in MATLAB with a time span of $t = [0, 1]$ seconds and a maximum step size of 0.01. For a given set of parameters, the differential equation was solved four times, one for each set of inputs. A multi-objective genetic algorithm in MATLAB using default settings was used to maximize the mean-squared difference of the output PC membrane potential across the different classes. The algorithm was allowed to change the connectivity parameters between columns, as well as the individual column parameters. The objective function also included a term to prevent the dynamics from diverging to infinity by penalizing the average mean-squared PC membrane potential of the output layer for times after $t = 0.5$ seconds.

\begin{figure}[b]
    \centering
    \includegraphics[width=0.9\linewidth]{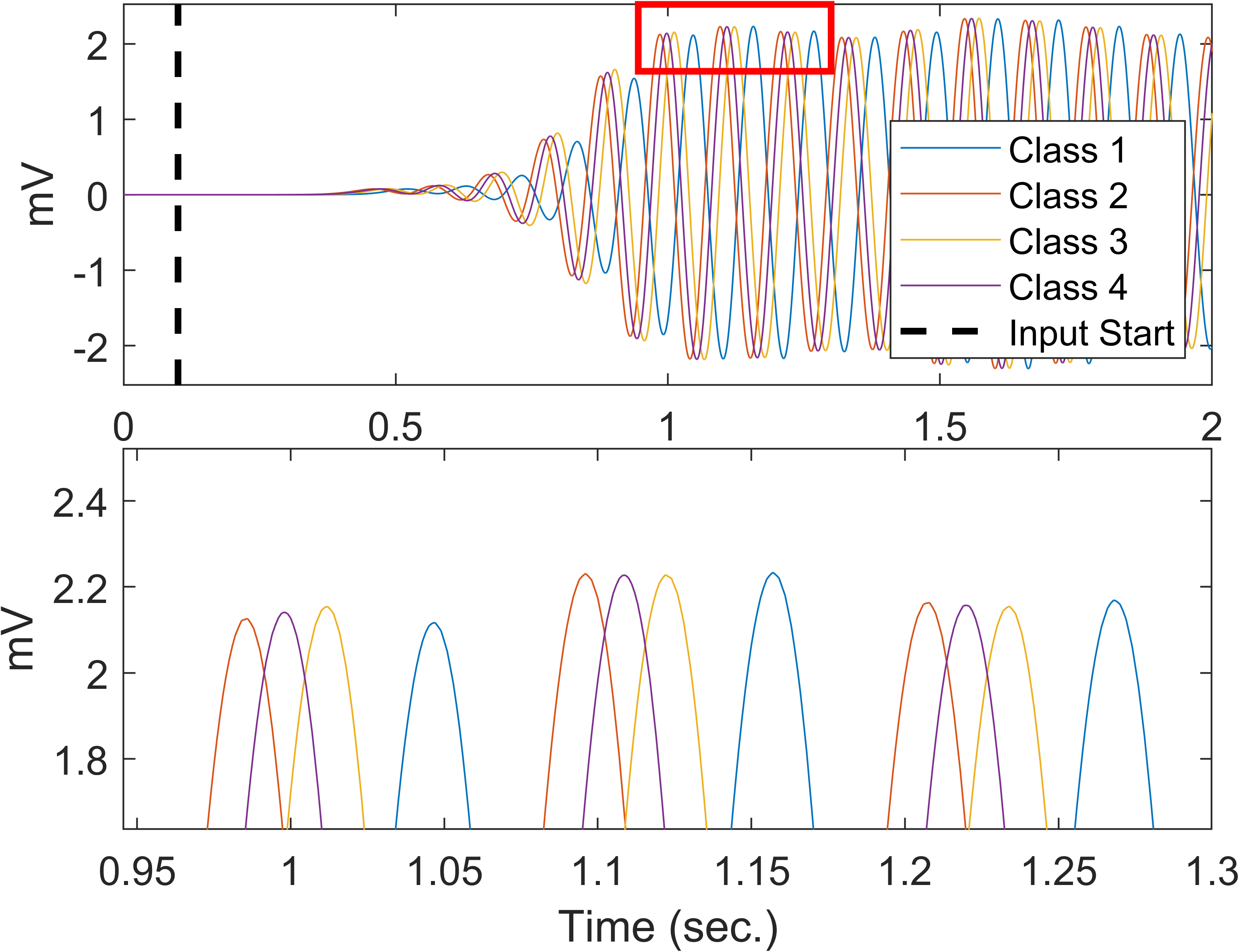}
    \caption{PC membrane potential of the output neuron for the different classes of inputs. The bottom plot is a zoomed-in section of the top plot.}
    \label{fig:timePhaseDiff}
\end{figure}

\section{Results}

\subsection{Phase Differences Between Classes}

\begin{figure}
    \centering
    \includegraphics[width=\linewidth]{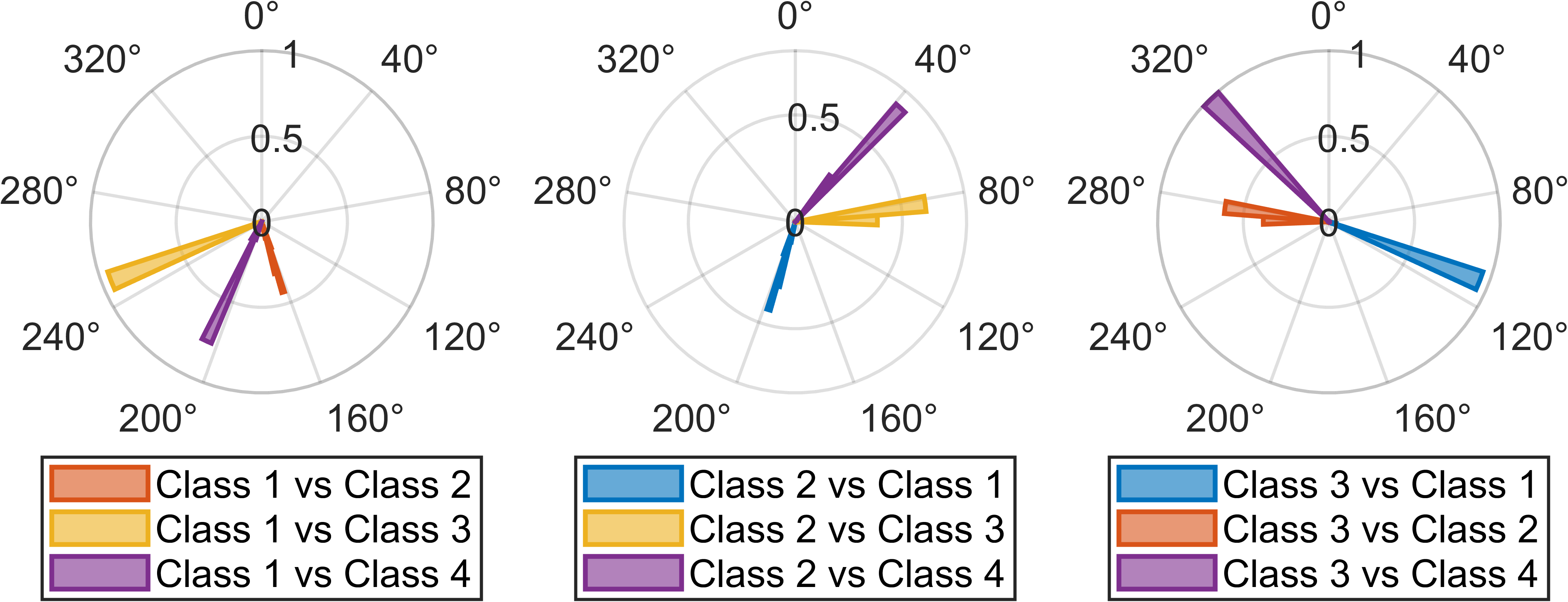}
    \caption{Pairwise phase distributions across the classes of inputs.}
    \label{fig:phaseDist}
\end{figure}

Fig.~\ref{fig:timePhaseDiff} shows the PC membrane potential of the output neuron across the different classes of inputs, after the genetic algorithm maximized the differences between classes. Qualitatively, the PC oscillates at approximately 8--10 Hz. Interestingly, the genetic algorithm maximized differences between the classes by finding parameters that led to phase shifts between the inputs. To quantify this observation, the distribution of the instantaneous phases between the classes is shown in Fig.~\ref{fig:phaseDist}. For each pair of classes, a Watson-Williams test indicated a significant phase difference with a p-value less than 0.05.

\subsection{Decoding of Phase into Oscillatory Power}

\begin{figure}
    \centering
    \includegraphics[width=0.9\linewidth]{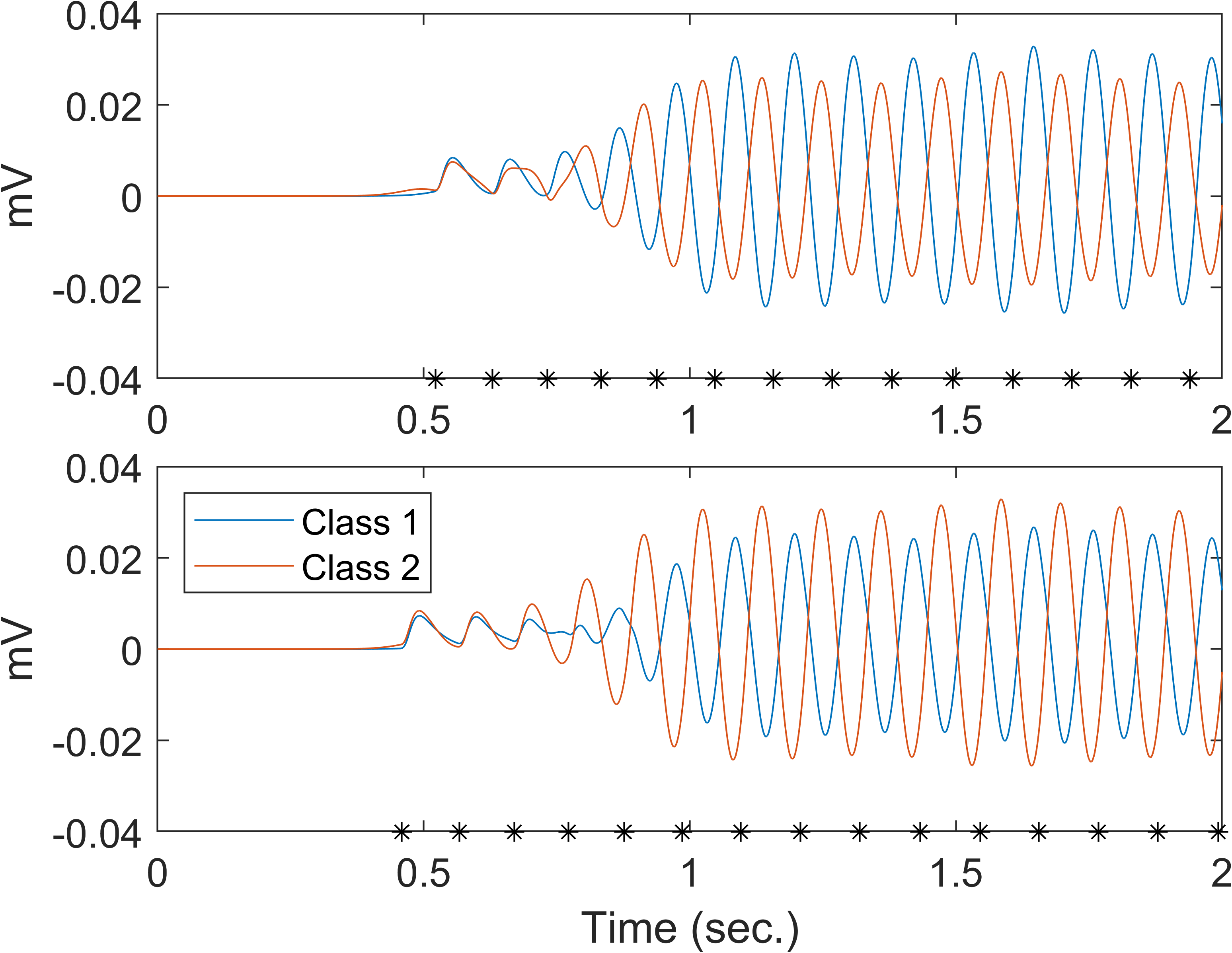}
    \caption{PC membrane potential of the output layer with peak-aligned pulse inputs. Inputs aligned with (top) Class 1 peaks and (bottom) Class 2 peaks. The black stars indicate the time the inputs were delivered.}
    \label{fig:figdecode}
\end{figure}

In this section, the phase encoding of the classes is decoded into oscillatory power. The output PC membrane potential from the previous section was taken for Class 1 and Class 2. Using the \texttt{findpeaks} function in MATLAB, the time points of the peak PC membrane potentials were extracted over time for Class 1. Pulses of inputs with amplitude 1 AP/s and duration 0.01 seconds were delivered to the EI layer of the output PC for both Class 1 and Class 2 inputs. When the pulses are aligned with the phases of the respective class, the oscillatory power is stronger, as shown in Fig.~\ref{fig:figdecode}.

\section{Discussion}

In this work, a biologically inspired neural network constructed using a JRNMM was optimized to maximize differences in neural outputs for a set of inputs. A genetic algorithm found parameters that resulted in phase shifts between oscillations for the sets of inputs, suggesting a method of encoding via phase. The phase shifts could be ``re-encoded'' or represented as changes in oscillatory power by aligning input pulses with the oscillatory peaks of a particular class. These findings provide insight into how a simple computational model of neural circuits can encode information through oscillatory dynamics.

To bridge the gap between traditional feed-forward networks and dynamical models of the brain, we constructed a Jansen and Rit Neural Mass Model (JRNMM) architecture comprising four input units, two hidden layers, and an output unit. Using a genetic algorithm, we optimized the network parameters (e.g., synaptic time constants, synaptic gains) to maximize differences in the output unit across four distinct classes through phase shifts in neural oscillations. Notably, this differentiation was achieved even though the inputs were delivered simultaneously across all classes. The phase encoding observed in our model parallels the theta phase precession seen in hippocampal place cells, where the firing phase relative to ongoing theta oscillations encodes an animal's spatial position \cite{o1993phase}. Similarly, in working memory, specific items are associated with particular phases of theta oscillations, enhancing the organization of multiple objects in real time \cite{siegel2009phase}. However, while these biological examples involve phase facilitating encoding through complex neural mechanisms, our model explicitly uses oscillation phases to encode distinct objects, demonstrating a direct computational role for phase in information representation. Future work will include exploring the dynamical system's parameters to explain how the nature of phase encoding arises in the model.

After demonstrating that the model is capable of encoding via phase, we explored a mechanism to ``decode'' the phases into another functional form (oscillatory power). Pulses of excitation were delivered to the output neuron at the peaks of membrane potential, in phase with a particular class response. When the pulses were in phase, the oscillatory power was stronger for that class compared to the others. This notion of oscillatory phase impacting neuronal dynamics has been observed in a variety of experiments. Low-frequency delta and theta oscillations have been shown to impact the responses to sensory stimuli by creating windows of high and low excitability \cite{schroeder2009low}. The change in oscillatory power could be used to modulate processing in another cortical region as a means of cortical-cortical communication via oscillations \cite{bonnefond2017communication}. Interestingly, nested oscillations were observed in the model presented, but analysis in this domain is left for future work.

\section{Conclusion}

In this work, parameters of a Jansen and Rit neural mass model were optimized using a genetic algorithm to maximize the model's response to distinct classes of inputs to explore computational models of neural encoding. The model that maximized responses encoded the different inputs as phase-shifted oscillations, which were decoded into oscillatory power by peak-aligning excitatory inputs. This provides a biologically inspired computational model for simple encoding-decoding in a neural circuit.

% Bibliography
\bibliographystyle{IEEEtran}
\bibliography{references} % references.bib file for citations

\end{document}